\documentclass[11pt]{aastex63} 
\usepackage{graphicx, natbib, amsmath}
\received{\today}
\revised{\today}
\accepted{later}
\submitjournal{ApJ}

\def\etal{{\it et al. }}

\def\msolar{\ifmmode{M_\odot}
    \else{$M_\odot$}\fi}
\def\lsolar{\ifmmode{L_\odot}
    \else{$L_\odot$}\fi}
\def\msun{\ifmmode{M_\odot}
    \else{$M_\odot$}\fi}
\def\lsun{\ifmmode{L_\odot}
    \else{$L_\odot$}\fi}
\def\ltsima{$\; \buildrel < \over \sim \;$}
\def\kms {km s$^{-1}$}
\def\cmcubed {cm$^{-3}$}
\def\vlsrk{V$_{LSRK}$}
\def\simlt{\lower.5ex\hbox{\ltsima}}
\def\gtsima{$\; \buildrel > \over \sim \;$}
\def\simgt{\lower.5ex\hbox{\gtsima}}

\def\H2{H$_2$}
\def\h3{H$_3^+$}

\shorttitle{Crab SNR Molecules}
\shortauthors{Wootten et al.}
\begin{document} 
\title{Dense Molecular Clouds in the Crab Supernova  Remnant}

\correspondingauthor{Alwyn Wootten}
\email{awootten@nrao.edu}

\author{Alwyn Wootten}
\affiliation{National Radio Astronomy Observatory, Charlottesville, VA, USA\footnote{The National Radio Astronomy Observatory  is operated by Associated Universities, Inc., under cooperative agreement  with the National Science Foundation.}}

\author{Rory O. Bentley}
\affiliation{UCLA Department of Physics and Astronomy, Los Angeles, CA 90095-1547, USA}

\author{J. Baldwin} 
\affiliation{Physics and Astronomy Department, Michigan State University, East Lansing, MI 48864-1116, USA}

\author{F. Combes} 
\affiliation{Observatoire de Paris, LERMA, Coll\`ege de France, PSL Univ., CNRS, Sorbonne Univ., Paris, France} 

\author{A. C. Fabian}
\affiliation{Institute of Astronomy, University of Cambridge, Madingley Road, Cambridge CB3 0HA, UK}

\author{G. J. Ferland}
\affiliation{Department of Physics and Astronomy, University of Kentucky, Lexington, KY 40506, USA}

\author{E. Loh} 
\affiliation{Physics and Astronomy Department, Michigan State University, East Lansing, MI 48864-1116, USA}

\author{P. Salome} 
\affiliation{Observatoire de Paris, LERMA, PSL Univ., CNRS, Sorbonne Univ., Paris, France} 

\author{C.N. Shingledecker}
\affiliation{Center for Astrochemical Studies, Max Planck Institute for Extraterrestrial Physics, Garching, Germany}
\affiliation{Institute for Theoretical Chemistry, University of Stuttgart, Stuttgart, Germany}
\affiliation{Department of Physics \& Astronomy, Benedictine College, Atchison, KS 66002, USA}

\author{A. Castro-Carrizo}
\affiliation{Institut de Radioastronomie Millimétrique, 300 rue de la Piscine, 38406 Saint-Martin-d’Hères France}


\begin{abstract}
Molecular emission was imaged with ALMA from numerous components near and within bright \H2-emitting knots and absorbing dust globules in the Crab Nebula. These observations
provide a critical test of how energetic photons and particles produced in a young supernova remnant interact with gas, cleanly
differentiating between competing models. The four fields targeted show contrasting properties but within them seventeen distinct molecular clouds are identified with CO emission; a few also show emission from HCO$^+$, SiO and/or SO. These observations are compared with Cloudy models of these knots.  It has been suggested that the Crab filaments
present an exotic environment in which H$_2$ emission comes from a mostly-neutral zone probably heated by cosmic rays produced in
the supernova surrounding a cool core of molecular gas. Our model is consistent with the observed CO J=3-2 line strength. These molecular line emitting knots in the Crab present a novel phase of the ISM representative of many important astrophysical environments.

\end{abstract}


\section{Introduction}

Massive stars explode via core collapse and ejection of surrounding layers. Large abundances of refractory elements in those layers may produce dust and molecules; these supernovae may be the dominant source of dust and molecules in the early Universe \citep{Sarangi+18}. ALMA has identified CO, $^{28}$SiO and $^{29}$SiO (\citet{Kamenetzky13}, \citet{Matsuura+17}) as well as copious amounts of dust ($\sim$0.7 \msun) in the core collapse supernova 1987A inner ejecta; the mass of molecular material continues to increase as the remnant evolves.  Dust (0.1-0.6 \msun\,  \citet{Barlow+10}, \citet{deLooze2017}) and molecules \citep{Walstrom+13} have also been observed in the more evolved (330 year-old) remnant Cas A, the remnant of a type IIb supernova of a massive supergiant.  Dust has also been measured in G54.1+0.3, which appears to have at least $\sim$0.3 \msun in the more evolved (1500-3000 year old) remnant of its 16 to 27 \msun progenitor \citep{Rho+18, Temim+17}.  Here we investigate the chemistry of a  remnant of intermediate age, the Tau A remnant, the Crab Nebula, which appeared in 1054.

The Crab Nebula is a prime example of a nearby and easily observable though more evolved young remnant of the common Type IIp core-collapse supernova. These SNe ejecta are efficient and productive dust sources among supernovae \citep{Sarangi+18}.  The Crab is a pulsar wind nebula well known for its expanding bubble of relativistic plasma that produces strong synchrotron radiation and cosmic rays. These in turn impinge on and excite a filamentary system of condensations that have formed around the exterior of the bubble. The ionized gas in these condensations has been heavily studied with special interest in the unusual chemical abundances produced in the SN.  The molecular content of the filaments has been little explored since the discovery of relatively bright infrared lines of H$_2$ (Graham et al. 1990).  Measurement of the molecular content can reveal gas isotopic composition, as well as details of the chemical evolution of the gas in the thousand years since the supernova.  The gas chemistry traces interaction of high-energy particles and photons with the molecular gas and may inform us how gas behaves in similar environments found in distant cool-core galaxy clusters, AGN and in SNe found early in the history of galaxies.

Data have been obtained from a detailed, pan-chromatic examination of the Crab's molecular component. 
We used our guaranteed time on the 4m SOAR Telescope to discover 55 individual knots that are 
strong sources of H$_2$ 2.12 $\mu$m emission (Loh et al. 2010, 2011). 
Our follow-up near-IR spectroscopy (Loh et al. 2012) showed that the H$_2$ excitation temperature 
is $\approx$3000 K. 
We assembled all pertinent HST narrow-band 
emission-line images as well as Spitzer and Herschel data (see examples in Fig 1). 
These show that the H$_2$ knots generally follow the optical filament system, but are 
associated with strong low-ionization optical emission lines from which we have measured the knots' velocities.

These data constrain detailed Cloudy models \citep{Richardson13}. The knots may be excited by synchrotron radiation, which is well constrained, or some other source of heating. 
The other sources of heating may be shocks or electrons and positrons from the pulsar. 
\citet{Priestley2017} find that a greatly
enhanced cosmic ray ionization rate, by 7 orders  of magnitude, over the standard interstellar value is required to account for the OH$^+$ and ArH$^+$ detected  lines, and lack of [C I] emission in the Herschel SPIRE FTS spectrum of the Crab  \citep{Barlow2013}.
The models that fit the K-band \H2 emission and optical lines show that the \H2 emission must come from what is really an extended HI zone in which a very high electron density leads to formation of \H2 through the H$^-$ route rather than in the usual way on dust. This zone is then heated, probably by cosmic rays produced in the Crab, to excite the trace amount of \H2. This is an exotic and almost completely unexplored environment for producing molecular emission, one that we seek to characterize with ALMA observations in this paper. 

To that end, we have also catalogued 280 small (arcsec scale) absorption blobs caused by dust silhouetted against the Crab's synchrotron continuum and compared their positions to the \H2-emitting knots. Many of these were recently discussed by \citet{Grenman+17}. A few \H2 knots are clearly associated with dust features, but most are not. However, the dust blobs often are associated with small knots of low-ionization gas for which we can measure velocities using our grid of optical spectra. The radial velocities show that most dust blobs are on the near side of the Crab's expanding shell, while most of the \H2 knots are on the far
side. Typical extinction through the dust blobs is A$_V \sim$ 0.5 mag, corresponding to N(HI) $\sim$ 10$^{21}$ cm$^{-2}$ up to a few times higher. This gas has so far not been seen in either 21 cm HI emission or absorption.  It would be confused with ISM emission in the LAB survey, and the published absorption maps (e.g. Greisen 1973) do not cover a sufficient velocity range.
Key questions include: What are the physical properties of the \H2-emitting knots, including the density, temperature, and mass? What is the physics that governs the molecular emission? What excites the molecular emission and what is its chemistry? What are the dust properties in this very young SN remnant? Is there also a component of cooler molecular gas? What is the connection between the dust blobs, the \H2 knots, and other molecular species such as CO? We address these issues with new data obtained with ALMA reported here.

\begin{figure}[ht!]
\label{fig:overview}
\plotone{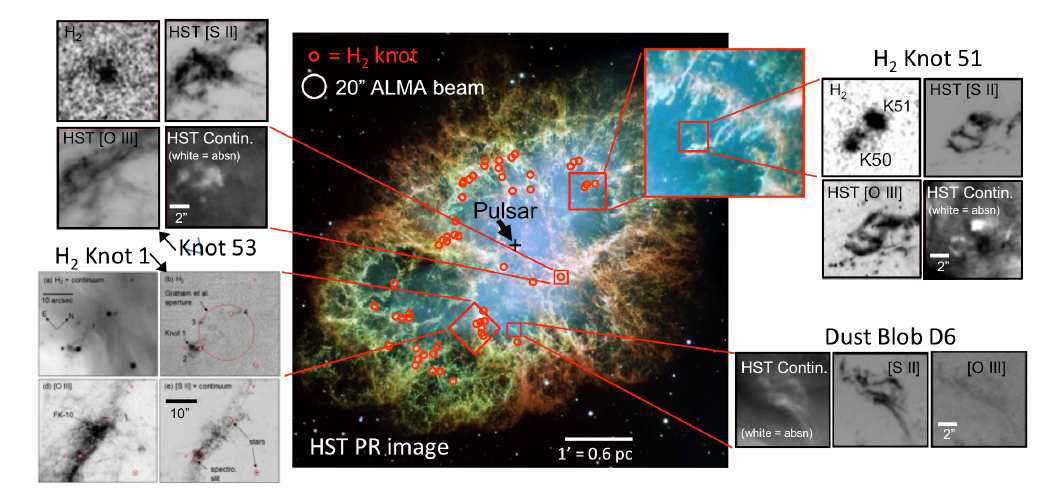}
\caption{An image of the Crab Nebula showing 55  \H2-emitting knots measured by \citet{Loh11} marked by red circles on the standard HST publicity image of the Crab.  Insets show enlargements of our four target positions in various ground- and space-based images.}
\end{figure}

We used ALMA to measure molecular emission from CO J=3-2, and from HCO$^+$, SiO and SO from four ALMA fields in the Crab nebula (Figure \ref{fig:overview}): (1) Knot 51, a bright \H2 knot that also has dust absorption; (2) Knot 1, a bright \H2 emitting knot that does not show dust absorption; (3) Knot 53, an \H2 knot for which we have observations seeking CO J=2-1 using the 30-m IRAM telescope and the Plateau de Bure interferometer; and (4) as a control sample, a dust blob D6 that is not a strong \H2 emitter. 


\section{Observations}

 Observations (Table~\ref{tbl:obs})
 were made with ALMA in the 850$\mu$m window (Band 7) in December 2015 on four occasions using a hybrid array during a transition from an extended configuration to a compact one. These data did not meet the brightness sensitivity proposed for the observations, therefore a second set of observations was obtained 16 June 2016 in a more compact configuration (C40-4) using 40 12m antennas deployed on baselines from 17 to 784m.  The largest scales to which the observations are sensitive are $\sim$11\arcsec.  The flux scale was set via observations of J0510+1800 (S$_{345.6GHz}$=1.08Jy). 
  
\begin{table}      [!htb]

\caption{Project: 2015.1.00188.S  Datasets Observational Summary}
\begin{center}
\begin{tabular}{lcccccc}
\noalign{\medskip}
\hline
\noalign{\smallskip}
  Date  & N$_{ant}$ & Baselines & PWV & Flux Calibrator & Flux\\
   &   &  m & mm &  & Jy\\
\hline
\noalign{\smallskip}
 12-Dec-15  &   31 &  15.2 - 7700 & 0.7 & J0237+2848 & 1.48\\
 12-Dec-15  &    31  & 15.2 - 7700 &  0.65 & J0510+1800 & 2.6\\
 15-Dec-15  &    43 & 15.1 - 6300 & 0.94 & J0423-0120  & 0.62 \\
 15-Dec-15 &    42 & 15.1 - 6300 & 0.94 & J0510+1800 & 2.6\\
 16-Jun-16  &    40 & 16.7 - 783.5 & 0.81 & J0510+1800 & 1.08\\
\hline
\hline
\end{tabular}
\end{center}
\label{tbl:obs}
Note: All integration periods were 2922 $\pm$ 4 s.
\end{table}
  
 Four pointings toward particular dust and/or H$_2$-emitting globules were made within a single session on 16 June.  In each pointing, four spectral  windows 1.875 GHz wide  were observed centered near 345 GHz to cover the four spectral lines listed in Table~\ref{tbl:spectro} to provide maximal continuum sensitivity and spectral grasp.  The raw data were processed by the ALMA/NA Regional Center using the CASA package 4.5.3; nearby quasars were used for flux and bandpass calibration as listed in Table  \ref{tbl:obs}. The phase calibrator used was J0521+2112 (S$_{345.6GHz}$=200$\pm$2.4mJy).

CASA clean was used in the delivered data with standard robust=0.5 weighting, tapered to 0.45'' beam to produce data cubes matching the proposal request for locating emission lines within the 17\arcsec  ~FWHP primary beam. All 1906 channels were imaged for each spectral window.  The beam size of the delivered data was 0.49 by 0.43\arcsec ~at position angle -37$^o$, with velocity resolutions listed in Table~\ref{tbl:spectro}.  In the delivered 16 June data, the sensitivity was 3.7mJy in those bandwidths.  Continuum sensitivity was about 75 $\mu$Jy.  The data was re-imaged to take advantage of the higher resolution which could be realized from all the data delivered (including Dec 2015 data) using natural weighting, which produced a 0.33 x 0.25\arcsec ~beam at PA -36$^o$ at a sensitivity of 2.8 mJy/beam or about 400mK.  It is on these images which we report.

\begin{table}
\begin{center}
\caption{Spectroscopic line parameters of in-band emission lines of interest$^a$.}
\begin{tabular}{ccccc}
\hline\hline
Species & Frequency&Transition &  E$\rm_{u}$  & $\delta $ v \\
  & (GHz) & & (K)& km s$^{-1}$ \\
 \hline \hline
CO                  & 345.79599 & 3-2         & 33.2   &  0.979 \\
SO                 & 346.528481 & 8(9) - 7(8) & 78.8& 0.979 \\
HCO$^+$$^b$      & 356.73422  & 4-3 & 42.8  & 0.979 \\
SiO                       &  347.33058& 8-7 & 75   & 0.979 \\ 
\hline
\end{tabular}

\tablenotetext{b}{HCO$^+$ J=4-3 at V$_{LSRK}>$400 \kms \ fell outside the tuning band.}

\end{center}
\label{tbl:spectro}
\end{table}

\section{Results}

Molecular knots in the Crab may be identified using different tracers.  The most complete description of the character of a knot includes aspects derived from all tracers.   For each knot, we discuss the relationship between the various tracer components-warm \H2, dust and target molecules-to develop a more comprehensive view of the character of the cool molecular emitting component.


\subsection{Knot 51}

\begin{figure}
\plotone{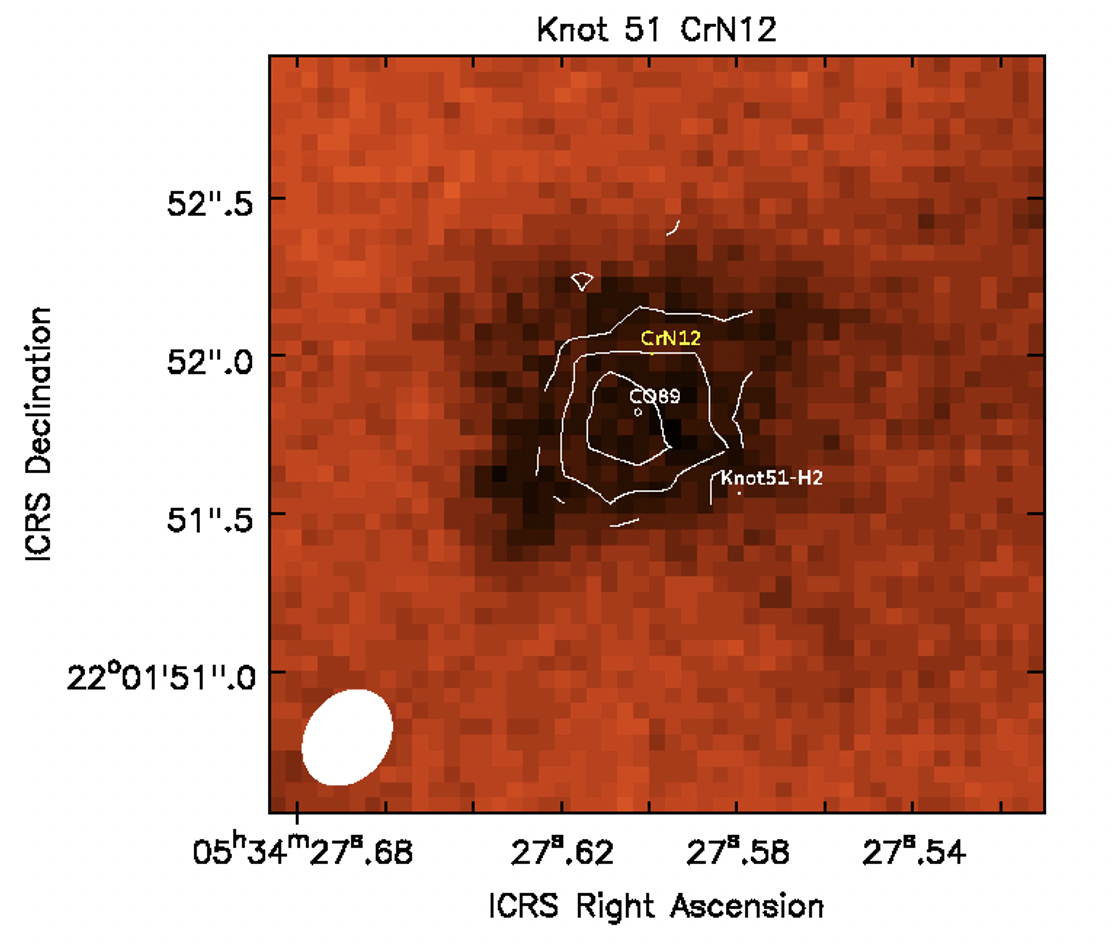}
\caption{\label{fig:k51}HST (observation GO-13510) image of the Knot 51 field with ALMA CO J=3-2 integrated intensity image (white contours) overlaid.  In the ALMA CO data, the rms is 2.8 mJy/bm \kms, where the beam is 0.33\arcsec $\times$ 0.26\arcsec.
Contours for CO emission are shown at 5 $\sigma$ intervals, at 5, 10, 15 $\sigma$.   CrN locates dust globule position in \citet{Grenman+17}; \H2 Knot51 lies at position Knot51-H2. \citep{Loh11}. }
\end{figure}

\begin{figure}
\plotone{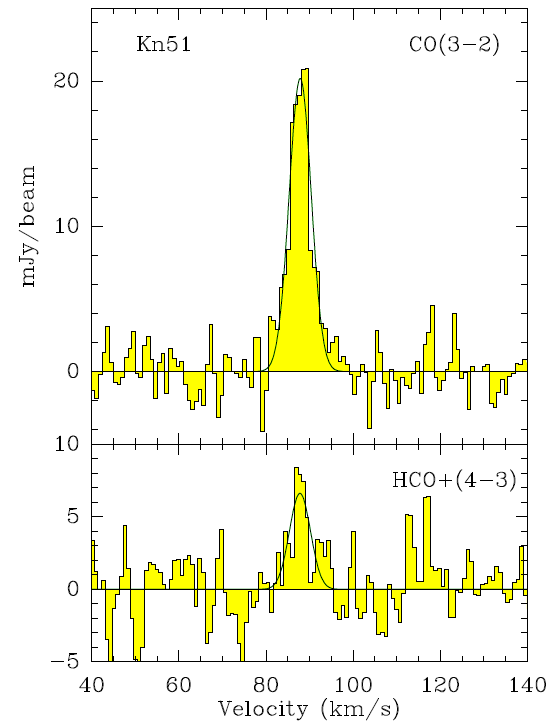}
\caption{\label{fig:k51sp}The spectra of CO J=3-2 (upper) and HCO$^+$ J=4-3 (lower) are shown for the position marked CO89 in the previous figure.}
\end{figure}

Spatially distinct and isolated, near the tip of an inwardly reaching finger of a gas filament to the northwest of the pulsar wind nebula, this $\sim1\arcsec$ diameter dust globule (CrN12 in Grenman et al 2017) was modeled by Richardson \etal 2013.  In this globule different tracers reveal different aspects of related spatial structures at different wavelengths.  \H2 traces higher excitation neutral gas.  A small \H2 radial velocity (V$_{LSRK}$=103.4 $\pm$ 100 \kms \citep{Loh12} coupled with the large transverse velocity (1340 \kms) measured for the dust knot CrN12 suggests that the knot lies near the nebular midplane, following its expansion westward in the plane of the sky, on the outskirts of the remnant \citep{Grenman+17}.   \citet{Grenman+17} measured CrN12 to subtend 1.2\arcsec (2400AU) diameter and estimated a dust mass for it of 1.6 $\times 10^{-5}$ \msolar. We clearly detect a distinct and 
 cospatial CO emission feature from Knot51, at V$_{LSRK}$=89 \kms~ illustrated in Figure~\ref{fig:k51}, suggesting association of the CO gas with both the dust and the warmer \H2 gas.
Allowing for nebular expansion of 0.01\arcsec/yr in the two years between the HST images of the dust knot and the ALMA observation, the integrated CO image closely follows the contours of the dust knot as seen in Figure 1, as does \H2.  This, along with the kinematic evidence, suggests close association between dust, \H2 and CO.  The full velocity width to half maximum intensity of the CO feature is 5 km/s; the deconvolved size of the CO emission is $0.46\arcsec \times 0.41\arcsec$ in a $0.33\arcsec\times0.26\arcsec$ beam.  HCO$^+$ emission was also seen from Knot 51 (Figure~\ref{fig:k51sp}) though only limits were measured for SiO and SO.   Knot 51 lies very close to Knot 50 and 52, both also sources of \H2 emission.  Knot 52 (dust counterpart CrN8 in \citet{Grenman+17}), with no CO counterpart, lies $\sim 5\arcsec$ to the west.  The third  \H2 knot, Knot 50, also without CO, lies $\sim2.5\arcsec$ southeast of Knot51 with the same radial velocity but its dust counterpart shows much less absorption (A$_v\sim 0.6-1.3$mag) compared to more than 2.6mag for Knot51 \citep{Richardson13}. No millimeter continuum emission was detected from the region, consistent with expectations given our sensitivity.


To try to understand the nature of Knot 51, 
\citet{Richardson13}
compared a number of different Cloudy models involving very different
density structures and excitation mechanisms.
In order to reproduce the observed H$_2$ intensity ratios, they found that
in addition to the Crab’s synchrotron radiation field, extra heating
by particles or some other source is required (their ``temperature floor''
and ``ionizing particle'' models). These models were arbitrarily truncated
at the back edge of the neutral zone that produces the H$_2$ emission,
and therefore predicted very little CO emission. 
 
However, as \citet{Richardson13} pointed out in their section 4.4,
there could also be an extensive zone of cooler molecular gas which
would produce strong CO and other molecular lines but no significant H$_2$.
As an example, they showed some results from a representative
``fully molecular core'' model which did not include the extra
heating sources needed to produce the observed H$_2$ spectrum,
but which (crucially) extended to a depth (5$\times 10^{17}$ cm) which was 8 times
greater than for the temperature floor and ionizing particles models.
Their figure 15 demonstrates the factor of $\sim 10^4$ times greater predicted
strength of CO lines (relative to the already-observed H$_2$  lines)
from the fully molecular core model as compared to the temperature floor model.

The CO observations described here require a great deal more molecular gas 
than is present in the temperature floor or ionizing particle models. 
The observed integrated CO 3-2 flux is 11 K km s$^{-1}$. The
temperature floor model predicts only 0.43 K km s$^{-1}$, 
which is about 25 times smaller than is observed,
while the ionizing particle model predicts $\sim 10^6$ times less 
CO emission than is observed. 
This shows that a significant fully molecular core must be present. 
However, the observed H$_2$ emission cannot come only from a cold
fully-molecular region, because the accompanying CO 3-2 emission would
then be far stronger than is observed.
Therefore, the true situation must be that the observed 
H$_2$ emission comes from a thin warm layer that emits little CO, 
while the CO emission comes from a moderately deep molecular core behind it.

  Our single observed transition of the rotational ladder does not provide strong constraints on the molecular excitation; 
  however the upper levels of the transitions detected in some knots (Table 1) lie between 33 and 79K suggesting a high excitation, 
  due to high volume density, as in the fully molecular core model. Then the CO excitation temperature should be equal to the dust temperature. 
  The \citet{Richardson13} model predicted grain temperatures between 38 and 54K, which they note is in agreement with \citet{Gomez12} 
  who fit unresolved Herschel submillimeter emission with cold and hot dust components at 28 and 63K respectively, 
  depending upon grain composition. We assume T$_{ex}$=40K in what follows, similar to the average dust temperature 
  for the nebula of $T_{dust}$=41$^{+3}_{-2}$K determined by \citet{DeLooze+19}. In the \citet{Richardson13} ``dense core'' model, 
  the density reaches n(\H2)=10$^6$  \cmcubed, sufficient to couple gas and dust temperatures, particularly in an environment 
  where excitation may involve electron as well as neutral particle collisions.  A neutral gas temperature of $\sim$ 40K and 
  density of $n \sim 10^6$~\cmcubed is also consistent with excitation estimates for SN87A \citep{Matsuura+17}; 
  the density is similar to that in the high-J CO knots in Cas A \citep{Walstrom+13}. Under these conditions, 
  the column density of CO in Knot51 is N(CO)=5$\times$10$^{15}$cm$^{-2}$ within the measured  dust core diameter of 1200 AU.  
  \citet{Grenman+17} estimated a dust mass for CrN12 of 1.6 $\times 10^{-5}$ \msolar .  
  Assuming a gas to dust ratio of 100 and a mean molecular mass of 2.33, we derive a CO abundance relative to  \H2 of {X(CO)=7.6 $\times 10^{-7}$}, somewhat lower than commonly cited ISM values \citep{Liszt2007}.
  Knot51 also shows HCO$^+$ emission; with the same excitation conditions we find an 
  HCO$^+$ column density of 
  N(HCO$^+$)=($5\pm .4$)$\times$10$^{12}$cm$^{-2}$ 
  and an abundance relative to \H2 of X(HCO$^+$)=7. $\times 10^{-10}$, using the LTE excitation modeling program  Radex \citep{vdT07+}.  Neither SiO nor SO emission were detected from this knot.  For all molecular knots, given the excitation estimates and 3$\sigma$ noise levels used here, column densities for undetected emission (S$\leq$3mJy/beam) are estimated to lie below N(HCO$^+$)=5$\times$10$^{11}$cm$^{-2}$, N(SiO)=3$\times$10$^{13}$cm$^{-2}$ and N(SO)=2$\times$10$^{13}$cm$^{-2}$.

\subsection{Knot 1}
\begin{figure}[h]
\centering

\plotone{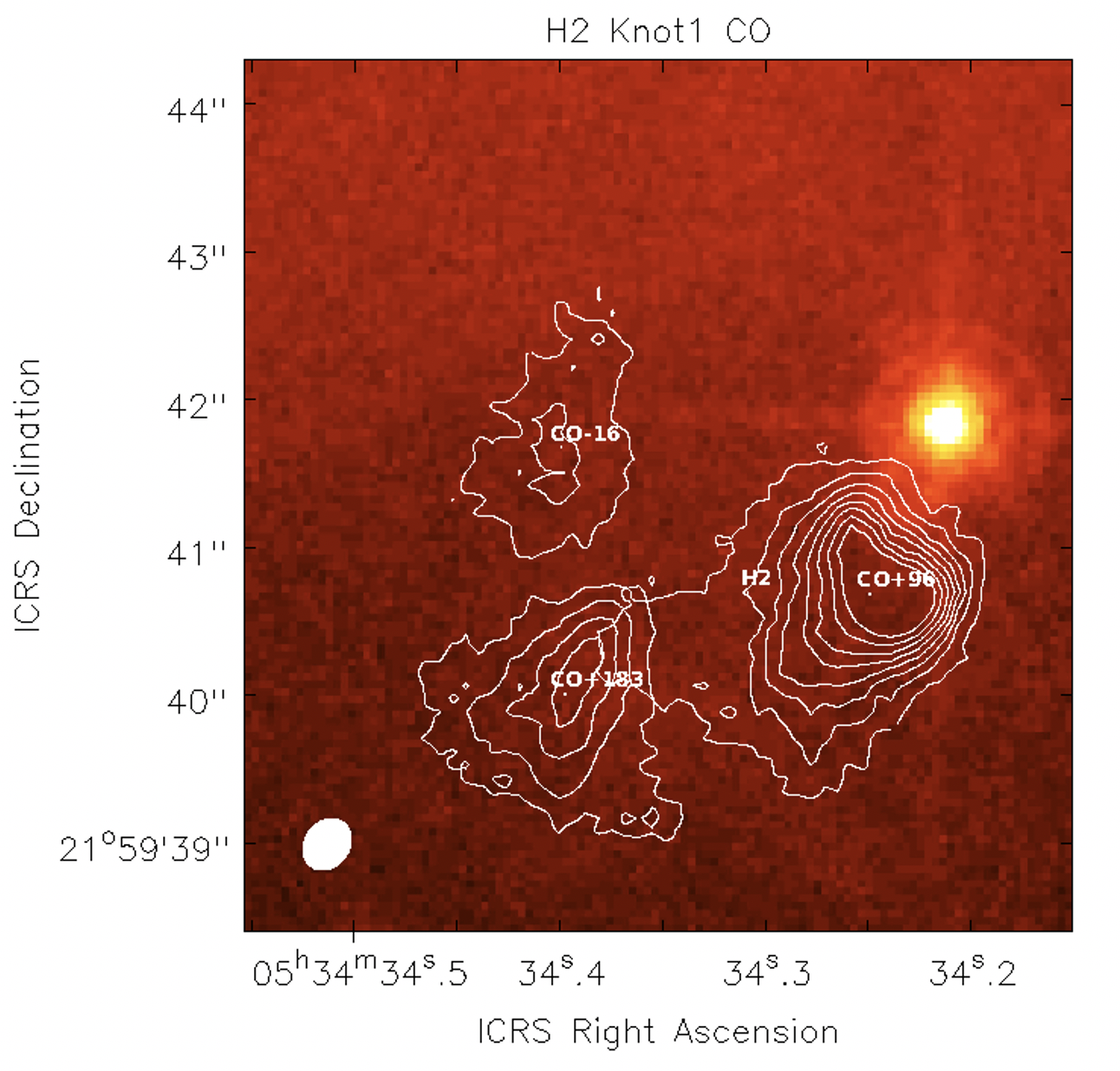}
\caption{\label{fig:f1mols}  HST (observation GO-13510) image of the Knot 1 field with ALMA moment 0 images of CO (contours) 
superposed.  Lowest contour is at 0.08 Jy/bm \kms and contour intervals are 0.16 Jy/bm \kms. 
  The weaker CO line at \vlsrk=-16 \kms has been hanning smoothed for this image. }
\end{figure}
Located directly on a very bright filament on the opposite (eastern) side but not so far ($\sim$ 0.6pc) in projection from the pulsar emission nebula as Knot 51, Knot 1 is a
complex, bright $H_{2}$ emitting knot \citep{Loh10} that does not have clear 
dust absorption in the Hubble continuum images. It is the brightest $H_{2}$ knot in the Loh catalogue \citep{Loh11} and among the largest, extending 5\arcsec $\times$ 2\arcsec \ approximately along an EW axis.  Knot 1 covers about four times the area of Knot 51 in \H2 emission.
The $H_{2}$ radial velocities (heliocentric) were measured to be 145 $\pm$ 100 \kms \citep{Loh12}.  The positive recession velocities suggest that from our viewpoint the material in Knot 1 lies on the far side of the remnant, illuminated on its face by the pulsar wind nebula, consistent with the absence of clear dust absorption.    Barlow \etal (2013) reported OH$^+$ and ArH$^+$ in this general vicinity (ArH$^+$ occurred at V$_{helio}$=140 $\pm$ 34 \kms, quite similar to the \H2 in the stronger Knot1 NW component).  

ALMA imaged three prominent distinct CO spatio-kinematic components in Knot 1 at radial velocities of \vlsrk=96 km/s, \vlsrk=183 \kms and \vlsrk=-16\kms as listed in Table~\ref{tab:results} and illustrated in Figure~\ref{fig:f1mols}. All components lay spatially and kinematically separated within a roughly 3\arcsec by 3\arcsec region in the center of the field, roughly centered on the location of Knot 1 in $H_{2}$, suggesting a physical association with that emission.  Line widths (FWHM) range up to  $\sim$40\kms, six times those observed in Knot 51.  Each component is distinct in its spatial and velocity characteristics suggesting that the three emitting regions lie separated along the line of sight within a complex structure.  The northwestern component is the stronger and consists of two intermingled sub-components oriented NE-SW with slightly more positive velocities to the SW; the variation is much less than the line width.  On the basis of position, CO integrated strength and velocity, we suggest Knot1NW-CO+96 is probably most closely associated with the $H_2$ cloud.
\citet{Loh12} estimated a lower limit to the $H_{2}$ molecular mass of 5$\times 10^{-5}$ $M_{\odot}$. 

We assume an excitation temperature for the molecules in Knot 1 as in Knot 51.  We estimate CO columns for the three molecular components as listed in Table~\ref{tab:results}, from 0.8 to 4 $\times$ 10$^{16} cm^{-2}$;  Knot1NW-CO+96 is the most prominent of these.  

Lacking dust absorption measurements, the total dust column in Knot 1 cannot be as well-characterized as it is in Knot 51.  \citet{Loh12} estimated that the total \H2 mass in Knot 1 could be 0.05\msolar, corresponding to a dust mass which might be detectable as thermal dust emission in our observations.  In fact, our slightly tapered continuum image ($0.71\arcsec \times 0.64\arcsec$ at PA $-29.50\deg$) for Knot 1 clearly showed continuum emission from the location of the \H2 over a region of similar extent to that subtended by the CO emission, at least for Knot1NW and Knot1E.
However, even using the more favorable gas to dust ratio of \citet{OwenBarlow15}  the continuum flux  translates to about 4 solar masses. This is unrealistic given the mass of the progenitor star.
It is likely dominated by synchrotron emission. Indeed, we detect for the two globules Knot1NW and Knot1E, over a surface of 40 beams, (or 3.6 10$^{-4}$ pc$^2$), a total integrated continuum emission S$_{870\mu m}$ = 25 mJy $\pm$ 12 $\mu$Jy.
The detected continuum is probably dominated by synchrotron emission by a factor $\sim$ 100, since the emission is too strong to be produced by dust.  Such non-thermal emission is expected since these globules belong to a conspicuous radio/optical filament.  Knot 1 appears to be the most massive observed knot, as suggested by its \H2 or CO line fluxes. It shows the brightest emission from each of the other molecules within the band, characterized in Table~\ref{tab:results}. Why are there continuum knots?
 ALMA observations of the 100 GHz continuum with a lower resolution \citep{Dubner2017} reveal multiple inhomogeneities, corresponding to  plasma confined to magnetic field lines, including wisps, arches and loops. 
 Lacking a clear radio counterpart or optical absorption, the dust mass of Knot 1 remains uncharacterized.

\begin{figure}[t]
\centering
\plotone{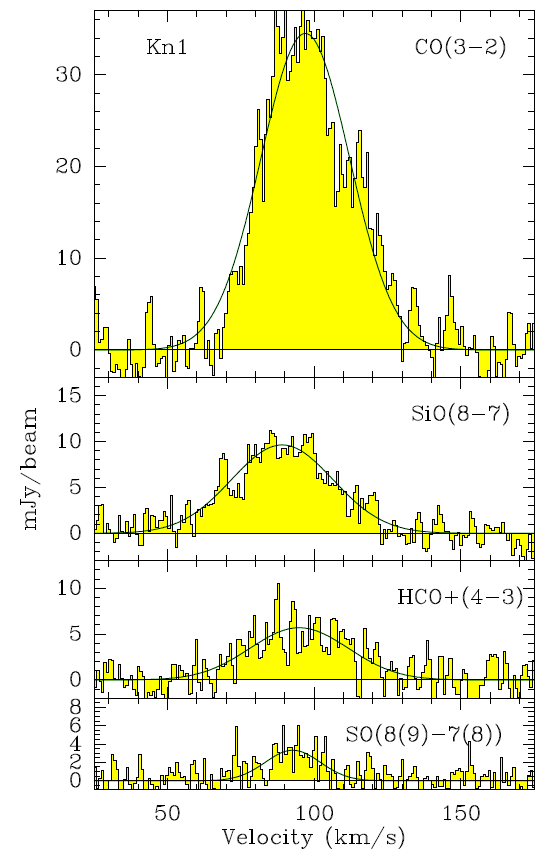}
\caption{\label{fig:f1profiles} Profiles of the strong 96 \kms lines in Knot 1: CO (3-2), SiO (8-7),  HCO$^+$(4-3) and SO(8(9)-7(8)).  }
\end{figure}

The 96 \kms  (NW) and the 183 \kms (E) components both showed emission from SiO J=8-7, SO 8(9) - 7(8) and HCO$^+$ J=4-3.   In Knot1-NW the SiO emission originates from the same region at the same velocity (Figure~\ref{fig:f1profiles}); however HCO$^+$ emission centers about 1000 AU NNE of CO and SiO.    Emission from all molecules is coincident in the E component.  Assuming the same excitation for all four molecules, we obtain the molecular column densities as listed in Table~\ref{tab:results}. 


\subsection{Knot 53}

\begin{figure}[t]
\centering
\plotone{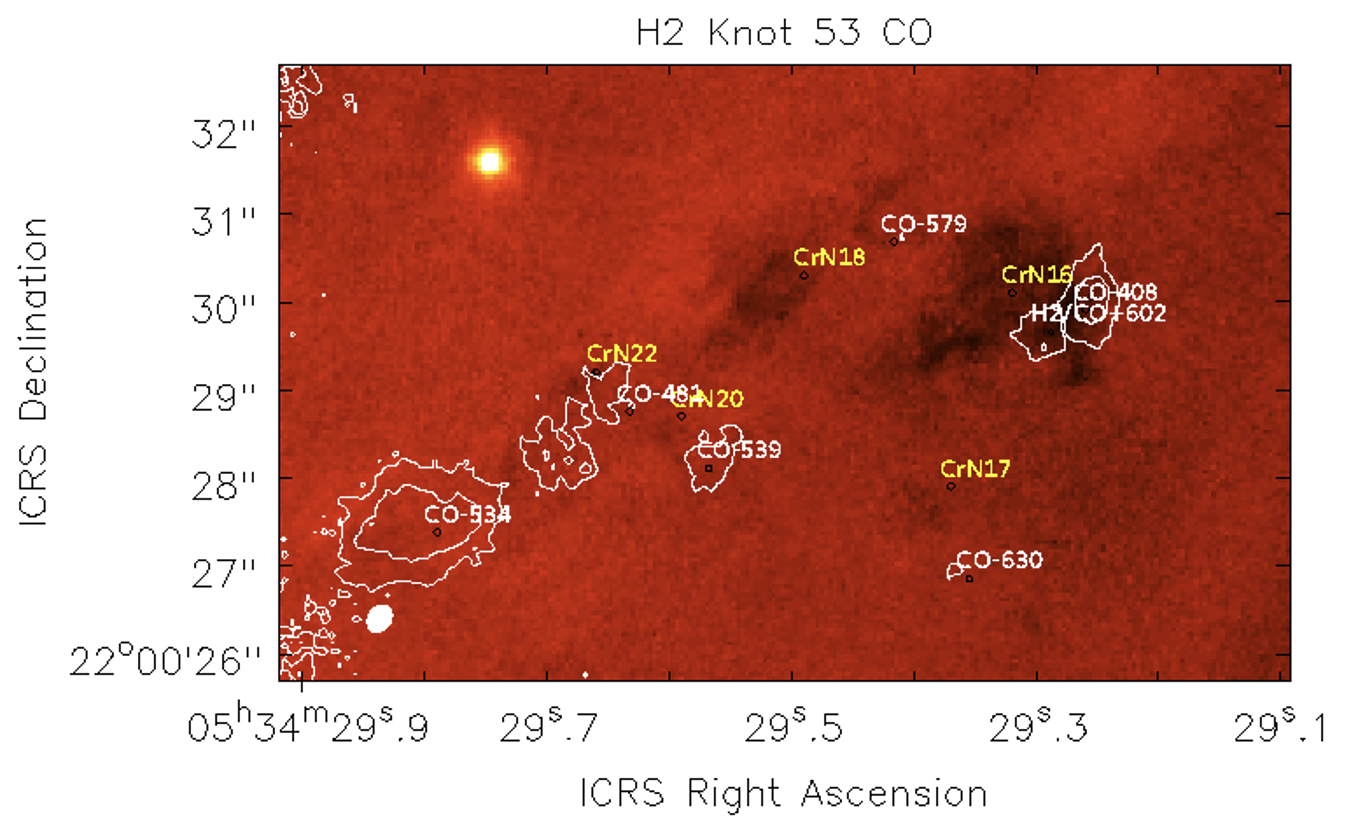}
\caption{\label{fig:Knot53}  (L) HST (observation GO-13510) image of the Knot 53 field stretched to emphasize the globules.  ALMA moment 0 images of the various CO J=3-2 velocity components are overlaid. CO- labels indicate different velocities.  White shows receding gas, in particular the 602\kms emission associated with the  \H2 knot 53; the nearby yellow shows the approaching gas at -408\kms emission associated with dust knot CrN16.    In the ALMA data, the rms is 2.8 mJy/bm \kms, where the beam is 0.33\arcsec $\times$ 0.26\arcsec; for that beam the rms is about 0.27K in brightness temperature.  Contours are shown at 45 $\sigma$ intervals.  CrN locates absorbing dust positions in \citet{Grenman+17}; \H2 Knot53 lies at position H2/CO+602. 
}
\end{figure}

\begin{figure}[h]
\centering
\plotone{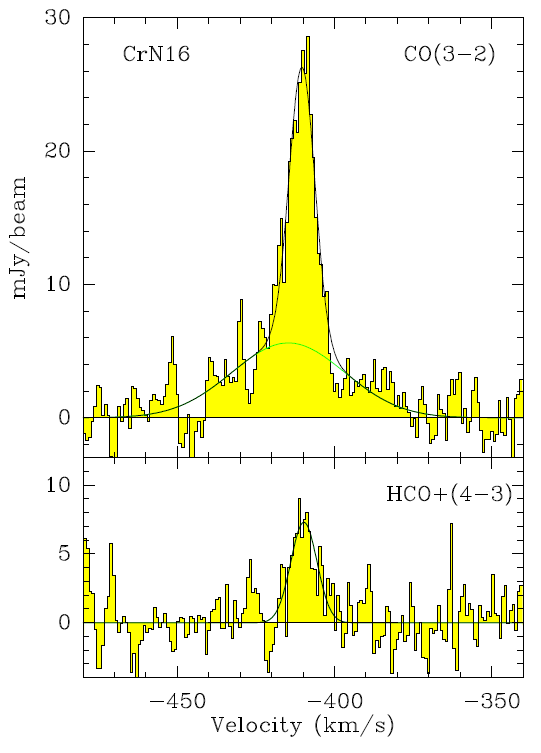}
\caption{\label{fig:Knot53sp} Profiles of CO J=3-2 and HCO$^+$ J-4-3 emission towards the CrN16 molecular emission peak.
The CO(3-2) spectrum is composed of a main component at -410 km/s (FWHM = 10. km/s, max = 21mJy)  which closely corresponds in extent with the dust extinction clump, and several nearby clumps without clear dust extinction counterparts, fitted by a wider gaussian, centered at -415 km/s, and of FWHM of 42 km/s (with a maximum at 5.6 mJy).
}
\end{figure}

Knot 53 is located closest in projection from the pulsar, and as might therefore be expected lies within a region of complexity and brightness exceeded only by the region around Knot 1. Knot 53 had previously been the target of CO J= 2-1 observations from the IRAM 30m telescope and the Plateau de Bure interferometer.
The noise level was rms 5 mJy in 2.5 km/s channels at 230GHz, 3 times higher than the ALMA rms at 345 GHz. The non-detection is compatible with the present ALMA results, since the expected CO(2-1) signal is lower by at least a factor 2 with respect to the  CO(3-2) signal (Salome et al, private communication).
This expected signal is assuming a dense molecular core, and a CO excitation temperature coupled to the dust temperature of 35-40K. This is consistent with the fully molecular core model of \citet{Richardson13}. Knot 53 is situated at the western head of a trail of dust knots seen in the HST image, which mark the loci of a major filament (`E' in the \citet{Hester+90} notation) in the Crab. The unresolved filament is also seen in Herschel images \citep{DeLooze+19}. There are several dense gas tracers identified within the ALMA primary beam, which targeted relatively strong molecular hydrogen emission at V$_{helio}$=696 \kms (V$_{LSRK}$=668 \kms).  Barlow \etal detected ArH$^+$ J=1-0 in detector C3 at V$_{helio}$=933 $\pm$ 33 \kms and ArH$^+$ J=2-1 in D4 at 743 $\pm$ 26 km/s in this direction.  \H2 and ArH$^+$ are probably associated with a particular far-side CO element, Knot53/H$_2$/+591.  CO emission in this region arises from two kinematic components, the molecular hydrogen emission component Knot53/H$_2$/+591 and an  CO feature at V$_{LSRK}$=-408 \kms, on opposite sides of the Crab at velocities separated 1000 \kms along one line of sight but less than half an arcsec in projected distance apart  (Figure \ref{fig:Knot53}).  CrN16 is a prominent dust globule coincident with the CO  feature, for which  \citet{Grenman+17}  estimated a dust mass of 5.8 $\times 10^{-5} \msun$  within a radius of 2\arcsec (2000AU).  \citet{Grenman+17} measured proper motion for the dust globules in this region; the expansion of the nebula moves them generally southwestward with time.  Between the \citet{Grenman+17} measurements in 2014 and the ALMA observations in mid-2016, all Knot53 features are expected to have moved 0\arcsec.14 generally along position angle 99$^o$, consistent with the differences we observe between dust and CO gas.   Positional coincidence suggests that the CO feature at -408\kms arises within CrN16. An HCO$^+$ emission peak also coincides with CrN16 in position and velocity (Figure \ref{fig:Knot53sp})

In all, ten CO features lie within the Knot53 field, with radial velocities between V$_{LSRK}$=-620.2 \kms and 601.5 \kms; seven are shown in Figure~\ref{fig:Knot53}, spread across a portion of the ALMA field of view.  Six of these show approaching velocities, centered east of Knot53-\H2, along the trail of dust features, spanning $\sim$5\arcsec \ with velocities between -630  and -408.0 \kms.   Knot53/CrN16/-408 lies at the western end of this chain.  

 We further identify CrN20 in Grenman's list as the host dark cloud for the CO emission feature at  V$_{LSRK}$=-539 \kms (W).   CO within a nearby distinct cloud to the east lies at V$_{LSRK}$=-534 \kms associating with very weak dust absorption uncatalogued by Grenman.  Dust globule CrN22 associates with CO at V$_{LSRK}$=-496 \kms .  Some prominent dark globules have no apparent CO emission--for example CrN18 is quite prominent but the only nearby CO, that at V$_{LSRK}$=-591 \kms, appears to be associated only with a small uncatalogued globule to its NE.~
 Even smaller CrN17 is, however, probably associated with weak broad CO emission near V$_{LSRK}$=-630 \kms.  
 
Three CO features with receding velocities lie in the far western portion of the field out of the Figure~\ref{fig:Knot53} field.  Knot53-CO+532 and Knot53-CO+540 lie west of Knot53-CO+286, oriented along PA 40 with V~530 in the NE increasing to 544 in the SW, broad and perhaps splitting in the SW.  These are SW of the main pulsar wind nebula with no clear associated features in the HST image.
 
  In summary, of the ten CO clouds we find in the Knot 53 field, six at positive velocities can reasonably be associated with near-side dust globules catalogued in  \citet{Grenman+17}, one cloud probably relates to the far-side \H2 and ArH$^+$ emission. Three at negative velocities seem unrelated to dust absorption in HST images or to H$_2$ emission.

Figure \ref{fig:Knot53sp} shows spectra of CO and HCO$^+$ towards the prominent CrN16  absorbing cloud.  We assume an excitation for the molecules in each of the Knot 53 clouds as in Knot 51.  As before, we use the  \citet{Grenman+17} dust mass and size estimates to derive a column density for \H2 which can then be compared to the CO column to provide an estimate of the CO abundance.  For instance, for CrN16,
we estimate N(\H2)=2.7 $\times 10^{22} cm^{-2}$  and thus derive a CO abundance relative to \H2 of X(CO)=7 $\times 10^{-7}$ , similar to CrN12 but lower than  commonly cited values for Galactic clouds \citep[e.g.,][]{Liszt2007}.
 
 
\subsection{Knot D6}
\begin{figure}[t]
\centering
\plotone{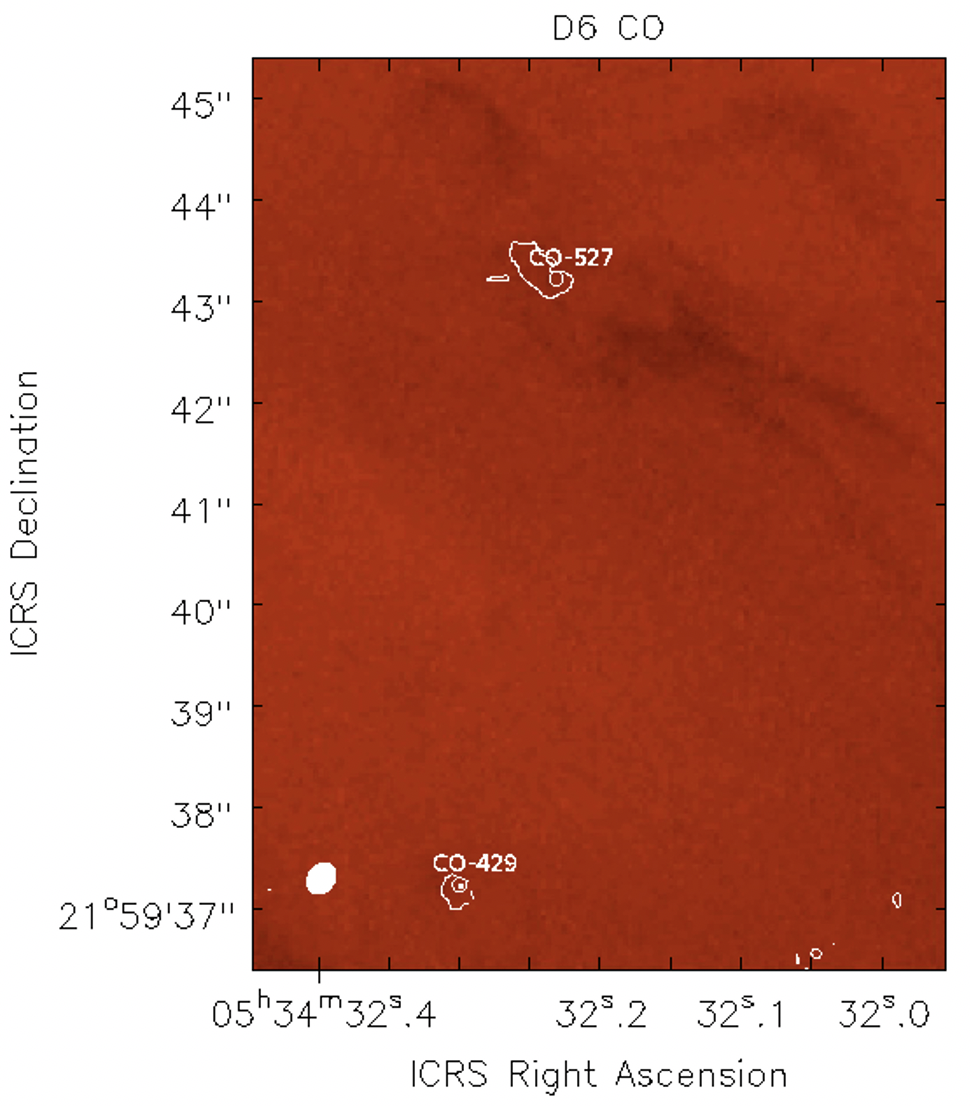}
\caption{\label{fig:KnotD6_center} HST (observation GO-13510) image of the D6 field with ALMA moment 0 images of two of the CO velocity components overlaid. Contours are as in Figure~\ref{fig:k51}.}
\end{figure}

 D6 lies 
 somewhat more distant ($\sim$0.8pc) in projection from the pulsar than other targeted features. \H2 was not detected toward D6 by Loh \etal (2011). Several dust features festoon the region, which includes features denoted as CrN 39, 40, 41 and 45 by \citet{Grenman+17}, with transverse velocities between 200 and 500 \kms.   None of the knots tabulated in that study coincides exactly with the dust feature D6 which sprawls around a small central CO feature, and which has a velocity of  V$_{LSRK}$=-521 \kms on the approaching side of the nebula (Figure \ref{fig:KnotD6_center}).  Barlow \etal detected ArH+ J=2-1 at V$_{helio}$-572 $\pm$ 25 km/s near D6, perhaps associated with the same feature.  
 In the eastern part of the field, the CO feature at  V$_{LSRK}$=-337 \kms lies about one synthesized beam southwest of the HST image location of dust globule CrN45 (Figure \ref{fig:KnotCrN45}). Both that dust feature and the CO feature are similar in size, $\sim$0.5\arcsec, slightly elongated in a SE-NW direction.  The offset of the centroids of the two features is consistent with the expected proper motion.  We identify CrN45 with the CO cloud on this basis. 
 A third CO feature was also detected south of the central D6 dust feature  at V$_{LSRK}$=-429 \kms,  but without apparent dust absorption in the HST image.  
 
   No emission was detected from HCO$^+$, SiO or SO toward any of the CO clouds in the D6 field.  We assume an excitation for the CO molecules in each of the clouds as in other knots.  As before, we use the  \citet{Grenman+17} dust mass and size estimates to derive a column density for \H2 which can then be compared to the CO column to provide an estimate of the CO abundance for the only cloud with CO and well-characterized dust, CrN45. There we estimate N(\H2)=4.1 $\times 10^{20} cm^{-2}$ and N(CO)=5.5 $\times 10^{15} cm^{-2}$ thus derive a CO abundance relative to \H2 of X(CO)=2 $\times 10^{-5}$, somewhat higher than in the previous knots where CO abundance could be derived.

\begin{figure}[t]
\centering
\plotone{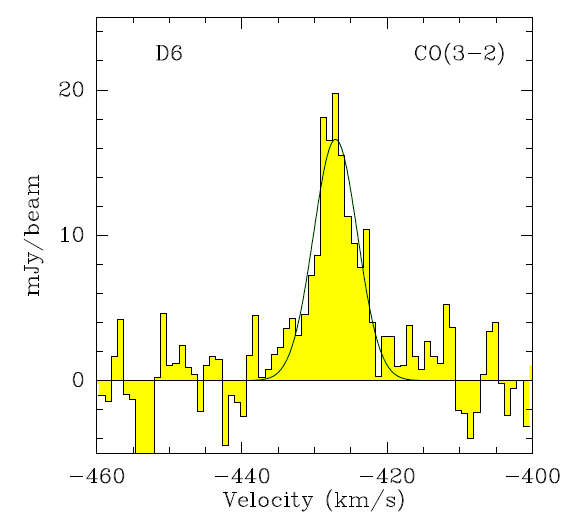}
\caption{\label{fig:KnotD6sp}  Profile of the V$_{LSRK}$=-427 /kms CO component toward D6.}
\end{figure}

 \begin{figure}[t]
\centering
\plotone{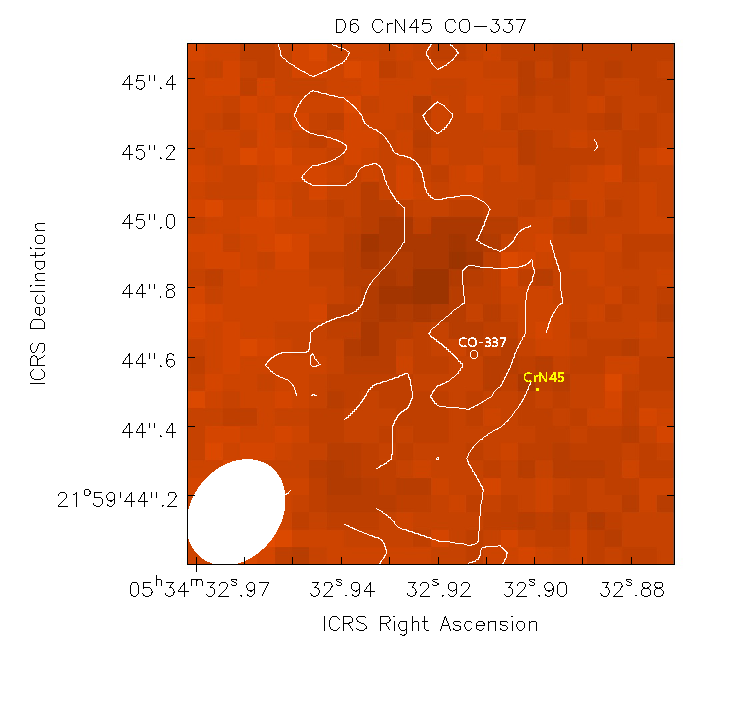}
\caption{\label{fig:KnotCrN45} HST (observation GO-13510) image of the CrN45 globule near the edge of our D6 pointing field with ALMA moment 0 images of the CO velocity component at V$_{LSRK}$=-337 \kms. Contours are as in Figure~\ref{fig:k51}.}
\end{figure}

\section{Discussion}

Perhaps the simplest of the fields is that of the isolated Knot 51, the subject of some previous modeling by \citet{Richardson13}.
The rough H$_2$-CO positional agreement does not suggest that the H$_2$ and CO emission come from a spatially identical region.
The most likely geometry is one where H$_2$ comes from very warm gas
near the surface of the globule, since temperatures approaching 3000 K are
needed to excite H$_2$ \citep{2008MNRAS.386L..72F},
while the CO originates in a much colder molecular core. 
The molecular hydrogen mass given in the last column of Table \ref{tab:results}
is the mass of the unobserved cold H$_2$ that is co-spatial with the CO. 
Their models required that the surface have an additional source of heating, which is almost certainly energetic particles \citep{Richardson13,Priestley2017}.
CO observations did not exist at that time so their comments on CO emission
are parenthetical and not the result of directed modeling.   Table \ref{tab:Abundances} provides estimates of CO, HCO$^+$ and SiO abundances for the seven globules where those molecules are observed which also have H$_2$ masses from Grenamn's dust measurements and where the extent of the molecular emission is the same as Grenman measured for dust.  

A variety of clouds are seen.  Seventeen globules have  CO emission, others have CO
and H$_2$, and seven have CO emission with associated dust.
Of the 17 CO emission regions we have described 
(Table 3), ten occur in regions of dust absorption.
These must occur on the near side of the nebula,
both to absorb the synchrotron continuum and to
account for that fact that nine have negative CO velocities. 
Four CO clouds also show \H2 emission, at positive velocities, but lack dust absorption.   
Since H$_2$ emission also occurs at positive velocities,  these clouds must lie on the far side of the expanding nebula.  
Four CO clouds have neither clear H$_2$ emission nor dust absorption. 

The distribution of CO emitters is symmetric in velocity suggesting that 
ALMA detects most of the clouds within the Crab in its limited field of view. 
The H$_2$ clouds are mostly receding \citep{Loh11}
suggesting that typically these have a large enough
column density to extinguish the 2 micron emission and hide the H$_2$  from our vision.
It follows that the extra heating needed to produce the H$_2$ emission must
predominantly deposit energy in the side of the globule facing the pulsar wind nebula.

The dust masses and column densities listed in 
Tables 3 and 4 have significant,  factors of $\sim 2$, uncertainties.
The Crab Nebula is chemically inhomogeneous with some regions
having an especially high helium abundance. The grain
material depends on the chemistry of the cold molecular
gas in which it forms and the grain size distribution,
which determines its reddening, depends on the grain
formation/destruction history. 
Despite this, the evidence suggests that the dust in the cores is similar to ISM grains.
Although there are models for each of these processes,
they are not robust and depend on many parameters.
\citet{Grenman+17} derived dust masses from Av measurements. The extinction law for dust grains in these globules matches a normal interstellar extinction law. This is evidence for a globule composition similar to normal interstellar grains.
We note that the ALMA observations detect
mainly the small globules and not the larger
filaments which have been studied by Herschel.
It would not be surprising if the dust in the
globules and filaments were different and this
introduces a further uncertainty in our analysis.
We provide these column densities from \citet{Grenman+17} for each absorbing globule to determine a point of reference; for Knot 1, with neither dust absorption (it is on the nebular far side) nor emission, only molecular column density is measured.

The fact that the Crab Nebula is young and not 
interacting with the ISM eliminates one major area of uncertainty, but also important is the fact that both the fluxes and energy distribution of locally generated
cosmic rays have likely not been significantly modified due to the interaction with ISM and can thus be
estimated fairly well \citep{Ivlev+18, Padovani+18}. The summary of molecular column density (\ref{tab:Abundances}) suggests an elevated HCO$^+$ abundance in Knot51 with respect to Knot 53, which may suggest changes in $\zeta$, but further observations are needed to secure this hint.

\section{Summary}
Spurred by \citet{Richardson13} who present a range of models for knots observed in the Crab, particularly the ``fully molecular core'' model of Knot 51, we have undertaken a small survey. We have imaged four 17\arcsec ~ fields of the Crab supernova remnant selected for \H2 emission and/or apparent dust absorption with ALMA, with spectral windows covering emission from CO, HCO$^+$, SiO and SO emitting in the 850 $\mu$m window.  Localized emission is found from seventeen clouds, typically under 1000AU in size, in the CO molecule.  Four knots show HCO$^+$ emission.  In H$_2$ Knot 1,  at both of the locations and velocities of two CO clouds all four molecules--CO, HCO$^+$, SiO and SO are detected.  These observations favor the ``fully molecular core'' model applying to a range of knot configurations.  
In this fully molecular core model, the emission of H$_2$, CO and other molecules does not  come from the same region. CO (and HCO$^+$, SiO, SO) emission occurs in the globule core, and H$_2$ on the surfaces. H$_2$ emission requires high temperatures suggesting cosmic ray excitation. Inside dense globules, the gas is cooler, H$_2$ may be present, but only the CO emission will be visible.

Our ALMA observations of Knot 51 show that a 
significant column density of molecular gas must be present, 
in addition to the thin skin that produces the H$_2$.
Figure 15 of \citet{Richardson13} shows that the
molecular emission produced by the fully molecular
core discussed in their paper is far more intense than their competing temperature floor model
that was optimized to reproduce only the H$_2$ observations. 
The H$_2$ emission originates in a warm skin so does not
require an extensive fully molecular core.
Our CO observations strongly support the fully molecular 
core for Knot 51 since a large column density of molecular gas is needed.

A new generation of spectral models is needed to really harvest the
information in the current data set or uniquely determine the energy source
of the molecular regions.  
The \citet{Richardson13} models focused on the remarkably strong H$_2$ emission.
Little was known of the CO spectrum so it was a parenthetical effort in that work.
As a result, these models were not optimized to determine
the energy source for the molecular core using the lines ALMA has detected. 
There has also been significant advances in the chemistry
(\cite{2021ApJ...908..138S, 2020RNAAS...4...78S} and in preparation)
and observations of molecules such as
ArH$^+$ are now available. 
A new unified modeling and observational effort may be able to
discern between such competing energy sources as
penetrating energetic particles or shocks.

The emitting regions encompass portions of a variety of structures observed at other wavelengths.  Three (\H2 Knots 1, 51 and 53) were selected as associated with known sources of molecular hydrogen emission.  CO was found to be associated with (but not cospatial with) \H2 in each of these; notably Knot 1 harbors three CO features, two of which also emit in the HCO$^+$, SiO and SO lines; all are less extensive than the warm H$_2$ emission. Of the latter two,  Knot1NW-CO+96 is the more massive but both may contribute to the H$_2$ cloud.  We note that a line survey of the much younger SN1987A remnant also showed emission from cool CO, HCO$^+$, SiO and SO molecules, possibly from post-explosion elemental mixing.  

Eight of the clouds showing CO emission, including Knot51, were found to be associated with absorption patches identified against the nebular emission as catalogued by \citet{Grenman+17}.  Four CO emission regions were not identified with absorption or with \H2 emission features.  Some \H2 emitting knots showed no associated CO (e.g. Knot 50) while some prominent absorption patches did not show detectable CO emission (e.g. CrN18). 

For the obscuring clouds \citet{Grenman+17} provided estimates  of visual absorption, 
which we have used to estimate the total column of \H2.  
Using assumptions for the excitation of CO, SiO, HCO$^+$ 
and SO based on \citet{Richardson13} models we estimated 
molecular column densities.  
The models do not predict substantial warm CO, 
consistent with  non-detection of higher-J CO lines 
in the much lower resolution Barlow et al (2013) SPIRE FTS spectra.
Abundances derived from the comparison of \H2 and molecular columns suggest that these globules  are chemically similar to the diffuse interstellar medium.
Therefore, in the thousand years since the supernova which produced the Crab Nebula, chemical processes have produced a molecular retinue similar to those seen in regions of presumably far more advanced age.  
For the few younger remnants for which there is data, a similar collection of molecules have been identified \citep{Matsuura+17}. 
Molecules can therefore provide good estimators of physical conditions within newly identified cool and dense clumps within the evolving remnant.  

Other regions, like the helium-rich belt, have very different compositions.
The Crab is probably also not chemically
homogeneous in the molecular knots that are the focus of this work.
Deep ALMA and JWST observations of these anomalous 
regions should be a high priority for future work.

The optical, IR, and now sub-millimeter observations suggest that the
globules have a common
geometry with a cool molecular core that is detected in CO,
surrounded by a warmer molecular layer emitting H$_2$.  
These are encased in an envelope of ionic emission seen in
the  HST images.
Individual knots are a combination of these basic ingredients,
with different amounts of each in different knots.
This shows that a surprisingly simple basic geometry can
account for what appears overly complex.

The \citet{Richardson13} study focused on understanding the
near infrared H$_2$ emission and was not guided by the
type of sub millimeter observations now possible with ALMA.
Nonetheless, they postulated that globules
with fully molecular cores were likely,
although they would be too cool to
contribute to the observed H$_2$ emission.
The CO and other sub-millimeter lines
we detect strongly support the existence
of such a fully molecular cores in 
some globules although it is not possible
to identify the heating mechanism. 
The low-J CO lines are thermalized so
mainly determine the temperature.
Other diagnostics, not available to the
\citet{Richardson13} study and so not
predicted by them, should
reveal the energy source for the 
molecular cores.
These should be the focus of future studies.

The geometry and spectra of Crab Nebula filaments are eerily similar
to the filaments seen in cool core clusters of galaxies
\citep{2008Natur.454..968F, 2009MNRAS.392.1475F, 2011MNRAS.417..172F}.
Indeed, this was the original motivation for the 
\citet{Richardson13} study, as they explain.
Even the gas pressures are analogous, with 
$ nT \sim 2\times 10^6$ cm$^{-3}$ K for Knot 51 in the Crab and 
$\sim 3\times 10^6$ cm$^{-3}$ K for well-studied regions
of the Perseus cluster. Only some differences might be noted:
the metallicity of the gas is higher in the Crab filaments, 
being expelled by the massive central star, while the cooling gas in
clusters is a mixture of primordial gas and ejecta from the central galaxy. 
The gas metallicity in the Perseus cluster is estimated to be 0.3 solar
\citep{Sanders2004}.
Dust is forming in supernovae ejecta \citep{Sarangi2018}, while it has difficulty
to survive the sputtering in clusters \citep{Draine1979}.
The similarities in the gas excitation provided the original
motivation to understand the far closer Crab Nebula filaments.
\citet{2009MNRAS.392.1475F} argued that the cluster filaments are excited
by energetic particles mixing with atomic and molecular gas.
Our current work shows that this process accounts for the emission 
from cool dense regions of Crab globules as well.  
The bright and well spatially resolved Crab Nebula thus provides a laboratory
to understand nearby analogues of the cool-core cluster environment.

\acknowledgments
The late Ed Loh played an important role in initiating this work.
We thank the referee for very helpful and detailed comments.
The National Radio Astronomy Observatory is a facility of the National Science Foundation operated under cooperative agreement by Associated Universities, Inc.  This paper makes use of the following ALMA data: ADS/JAO.ALMA2015.1.00188.S. ALMA is a partnership of ESO (representing its member states), NSF (USA) and NINS (Japan), together with NRC (Canada), MOST and
ASIAA (Taiwan), and KASI (Republic of Korea), in cooperation with the Republic of Chile. The Joint ALMA Observatory is operated by ESO, AUI/NRAO and NAOJ.  Rory Bentley was a summer student at the National Radio Astronomy Observatory.
GJF acknowledges support by NSF (1816537, 1910687), NASA (ATP 17-ATP17-0141, 19-ATP19-0188), and STScI (HST-AR- 15018 and HST-GO-16196.003-A).

\begin{table}[t]
\small
\begin{center}
\caption{Molecular Features in the Crab SNR. \footnote {H$_2$ masses have been derived, assuming  CO/H$_2$=1 10$^{-5}$, an average for the seven globules with both dust mass and CO column density measured.}}
\footnote{The globule properties were derived from observations reported here, along with H$_2$ and optical observations.  Values reported in this table are for beam averaged values at the position given. Column densities are derived using Radex, a molecular excitation fitting program, using the robust excitation (Tx=40K, density $\simeq$10$^6$ cm$^{-3}$  suggested by our detection of high-excitation transitions, and by \citet{Richardson13}.
}
\vskip 2mm 
\label{tab:results}
\begin{tabular}{lcccccccc}
\hline\hline
{Field}   & {RA} & {Dec} & Velocity & Width & {Int. Flux} & {FWHM}  &{N(X)}& M(H$_2$)\\
{ } & {J(2000.0)} &{J(2000.0)} & {\kms} & {\kms} &{K \kms} & {arcsec} & {cm$^{-2}$}&10$^{-4}$M$_\odot$\\
\hline
Knot51/H$_2$/CrN12/89 &5:34:27.603 & 22:01:51.81 & 89. & 5 & 11 & 0.4 & $(5.5 \pm 0.2)\cdot10$$^{15}$&  1.7\\ 


Knot53/CrN17/-630 &5:34:29.355 & 22:00:26.85 & -630 & 30. & 30 & 0.4 & $(1.5 \pm .05)\cdot10^{16}$& 6.7\\ 

Knot53/CrN18/-590 & 5:34:29.417 & 22:00:30.68 & -590 & 1.6 & 2.4 & 0.5 & $(1.2 \pm 0.5)\cdot10^{15}$&0.6\\ 

Knot53/CrN20W/-539 &   5:34:29.568 & 22:00:28.11 & -539 & 11.7 & 33 & 0.5 & $(1.6 \pm 0.7)\cdot10^{16}$ & 8.4\\

Knot53/CrN20E/-534 &   5:34:29.790 & 22:00:27.38 & -534 & 14.3 & 54 & 0.6 & $(2.8 \pm 1.3)\cdot10^{16}$ & 14.3\\

Knot53/CrN22/-496  & 5:34:29.633 & 22:00:28.75 & -496 & 8.1 & 25.9 & 0.6 & $(1.3 \pm 0.2)\cdot10^{16}$ & 6.1\\

Knot53/CrN16/-408 &   5:34:29.260 & 22:00:29.91 & -408 & 8 & 32. & 0.9 & $(1.7 \pm 0.1)\cdot10^{16}$ & 88.4\\

Knot53/+286 &   5:34:28.531 & 22:00:34.6 & 286 & 3 & 18 & 0.5 & $(9.5\pm 1.0)\cdot10^{15}$& 4.5\\

Knot53/+532 &   5:34:28.479 & 22:00:33.09 & 532 & 9 & 17.5 & 0.4 & $(2.9 \pm 1.4)\cdot10^{16}$ &  3.0\\

Knot53/+540 &   5:34:28.457 & 22:00:32.47 & 540 &11 & 70 & 0.4 & $(3.8 \pm 0.3)\cdot10^{16}$ & 8.4\\

Knot53/H$_2$/+591 &   5:34:29.288 & 22:00:29.65 & 591 & 5.6 & 21. & 0.8 & $(7. \pm 0.5)\cdot10^{15}$ &14.8\\ 

Knot1-CO-16    & 5:34:34.40 & 21:59:41.67 & -16 & 15 & 20  & 0.5 & $(9.0 \pm 0.3)\cdot10^{15}$ & 4.2\\ 

Knot1NW-CO+96  & 5:34:34.250 & 21:59:40.68 & 96 & 28 & 98  & 1.3 & $(5.0 \pm 0.3)\cdot10^{16}$&  142. \\ 

Knot1E-CO+183    & 5:34:34.398 & 21:59:40.00 & 183 & 15 & 45 & 1.8 & $(2.3 \pm 0.1)\cdot10^{16}$ & 143.\\ 

D6-CO-429  & 5:34:32.30 & 21:59:37.24 & -429 & 5 & 13 &0.3 & $(6.3 \pm 0.1)\cdot10^{15}$& 1.1\\

D6-CO-521  & 5:34:32.231 & 21:59:43.22 & -521 & 7 & 4.3 &0.3&$ (2.0 \pm 0.4)\cdot10^{15}$& 0.37\\

D6/CrN45-CO-337  & 5:34:32.913 & 21:59:44.58 & -337 & 4.5 & 15 & 1.1 &$ (7.5 \pm 0.2)\cdot10^{15}$& 17.8\\
\hline
Knot51/CrN12/-HCO$^+$~89 &  5:34:27.603 & 22:01:51.80 & 87.5 & 5 & 5.5 & 0.4  &$(5.0 \pm 0.4)\cdot10^{12}$ &\\ 

Knot53/CrN16-HCO$^+$-408 & 5:34:29.25 & 22:00:30.1 & -408 & 8.1& 8.1 &0.9  & $(8.0 \pm 0.4)\cdot10^{12}$ &\\ 

Knot1NW-HCO$^+$~+96 & 5:34:34.256ck& 21:59:40.99 & 94. & 40 &  50 & 0.8 &$ (4.5 \pm 0.1)\cdot10^{13}$& \\ 

Knot1E-HCO$^+$~+183& 5:34:34.387ck& 21:59:40.19 & 182 & 25 & 45 & 0.5 & $(4.4 \pm 0.1)\cdot10^{13}$& \\ 
\hline
Knot1NW-SiO~+96  & 5:34:34.256ck& 21:59:40.99 & 94. & 40 &  50 & 0.8 & $(2.0 \pm 0.1)\cdot10^{15}$& \\ 

Knot1E- SiO~+183  & 5:34:34.387ck& 21:59:40.19 & 182 & 25 & 45 & 0.5 & $(1.8 \pm 0.1)\cdot10^{15}$& \\ 

\hline
Knot1NW-SO~+96 8$_9$ - 7$_8$ & 5:34:34.256 & 21:59:40.99 & 96. & 20 &  13 & 0.8 & $(5.0 \pm 0.5)\cdot10^{14}$& \\ 

Knot1E-SO~+183 8$_9$ - 7$_8$ & 5:34:34.387 & 21:59:40.19 & 183 & 22 & 11 & 0.5 & $(9.0 \pm 0.1)\cdot10^{12}$& \\ 
\hline
\end{tabular}
\end{center}
\end{table}

\bigskip

\begin{table}[t]
\small

\begin{center}
\caption{Observed relative abundances in the Crab SNR Molecular Globules}
\vskip 2mm 
\label{tab:Abundances}
\begin{tabular}{rlccc}
\hline\hline
{Field} & {Transition Line} & {X Column Density} &  {\H2 Column Density} & {Abundance Estimate}\\
{ } & { } & {cm$^{-2}$} & {cm$^{-2}$} & {n(X)/nH}\\
\hline
Kn51/CrN12  & CO 3-2 &  (5.5$\pm$ .2) $\times$ 10$^{15}$ & 7.5 $\times$ 10$^{21}$ &   7. $\times$ 10$^{-7}$\\
Kn53/CrN16  & CO 3-2 &  (1.7$\pm$ .1) $\times$10$^{16}$  &  2.7 $\times$ 10$^{22}$ &   7. $\times$ 10$^{-7}$ \\
Kn53/CrN17  & CO 3-2 &  (1.5$\pm$ .1) $\times$10$^{16}$  &   1.4 $\times$ 10$^{21}$ &   1.1 $\times$ 10$^{-5}$ \\
Kn53/CrN18  & CO 3-2 &  (1.1$\pm$ .1) $\times$10$^{15}$  &   7.1 $\times$ 10$^{21}$ &   1.6 $\times$ 10$^{-7}$ \\
Kn53/CrN20W  & CO 3-2 &  (1.6$\pm$ .7) $\times$10$^{16}$  &   9.4 $\times$ 10$^{20}$ &   3.1 $\times$ 10$^{-5}$ \\
Kn53/CrN22  & CO 3-2 &  (1.3$\pm$ .2) $\times$10$^{16}$  &   7.1 $\times$ 10$^{21}$ &   1.6 $\times$ 10$^{-7}$ \\
Kn51/CrN45  & CO 3-2 &  (5.5$\pm$ .2) $\times$ 10$^{15}$ &  4.1 $\times$ 10$^{20}$ &   1.7 $\times$ 10$^{-5}$\\
Kn51/CrN12 & HCO$^+$ 4-3 &  (5 $\pm$ .4) $\times$ 10$^{12}$  &   7.5 $\times$ 10$^{21}$ &   7 $\times$ 10$^{-10}$ \\
Kn53/CrN16 & HCO$^+$ 4-3 &  (8$\pm$ .4) $\times$ 10$^{12}$ &  1.4 $\times$ 10$^{21}$ &  6 $\times$ 10$^{-9}$\\

\hline

\end{tabular}
\end{center}

\end{table}

\bibliography{Crabads}{}
\bibliographystyle{aasjournal}

\end{document}